\documentclass[conference]{IEEEtran}

\usepackage{epsfig,subfigure}
\usepackage{amssymb, amsmath}
\usepackage{color}

\usepackage{graphicx}

\usepackage{amssymb}
\usepackage{amsmath,amsfonts,amssymb}
\usepackage{verbatim}
\usepackage{stfloats}
\usepackage[bookmarks=false]{}

\newtheorem{theorem}{Theorem}

\begin{document}
%
%
\date{}
\title{Topological Interference Management with Alternating Connectivity}
\author{\IEEEauthorblockN{Hua Sun, Chunhua Geng and Syed A. Jafar}
\IEEEauthorblockA{University of California Irvine, Irvine, CA 92697\\
Email: \{huas2, chunhug, syed\}@uci.edu}}

\maketitle

\begin{abstract}
The topological interference management problem refers to the study of the capacity  of partially connected linear (wired and wireless) communication networks with no channel state information at the transmitters (no CSIT) beyond the network topology, i.e., a  knowledge of which channel coefficients are zero (weaker than the noise floor in the wireless case). While the problem is originally studied with fixed topology, in this work we explore the implications of varying connectivity, through a series of simple and conceptually representative examples. Specifically, we highlight the synergistic benefits of coding across alternating topologies.

\end{abstract}

\section{Introduction}
Network coding has blurred the dichotomy of wired and wireless communication networks. With network coding, and linear network coding in particular performed at the intermediate nodes, an end-to-end wired network behaves much the same way as a wireless interference network, albeit over finite fields or packets rather than real or complex signals. Interference is introduced by inter-session coding at the relay nodes, which transmit linear combinations of their incoming symbols on their outgoing links, thus creating an end-to-end linear interference network. It is then natural to try to transfer the emerging interference management principles from wireless to wired networks \cite{Jafar_tutorial, Jafar_TIM}.

In transferring insights from wireless networks to wired networks, limitations follow as well. Interference management schemes such as interference alignment for wireless networks often require abundant channel state information at transmitters (CSIT) which is analogous to learning the exact end-to-end coding coefficients in the wired case, which can constitute a significant overhead for larger field sizes. In both wired and wireless networks,  exact channel knowledge is difficult to obtain, which makes these theoretical insights difficult to translate into practice. Therefore there is much interest in studying settings with relaxed channel knowledge assumptions.

A complementary perspective, called topological interference management, is introduced in  \cite{Jafar_TIM} where the focus is on  minimal channel knowledge assumptions --- the transmitters are only aware of the network topology through 1-bit feedback indicating whether an interference link is present or not, but no knowledge of channel coefficient value is available to the transmitter. The topological interference management problem takes a unified view of both wireless networks and wired networks with linear coding at intermediate nodes, showing that the degrees of freedom (DoF) in the wireless case and the capacity in the wired case, are often the same value in their respective normalized units, determined by the same principles, so much so that a solution to one problem automatically solves the other. The study of the capacity in the wired case and the DoF in the wireless case, for  partially connected network with no CSIT except the network topology is called the topological interference management problem. When the channel coefficients are fixed  the topological interference management problem is shown in \cite{Jafar_TIM} to be essentially related to the index coding problem \cite{Yossef_Birk_Jayram_Kol_Trans, Hamed_index}. Applications to time-varying channels for a fixed topology are explored  for sufficiently large coherence intervals in \cite{Jafar_CBIA}, where also an example with coherence interval = 1 is presented to show the distinct character of the problem under insufficient coherence. The latter setting is studied extensively in \cite{Navid_Salman}. In all these studies however, even when the channels are assumed time-varying, the network topology is assumed to be fixed throughout the duration of communication.

The purpose of this work is to explore the problem as we go beyond this limitation. Network topology can vary  in a wired network as the linear network coding coefficients are varied, and  in a wireless network with  frequency-hopping or  multi-carrier transmission in frequency selective environments. So in this work we explore the topological interference management problem with time-varying (alternating) connectivity. We will focus on wired networks and exact capacity results for the exposition in this paper, but as shown in \cite{Jafar_TIM} the results can be directly translated into degrees-of-freedom (DoF) results for corresponding instances of wireless networks. 

\section{System model}
Except for the assumption of alternating topology, we will retain the framework of \cite{Jafar_TIM}. As such, consider the class of linear wired interference networks where the output at each receiver is a linear combination of the inputs from the transmitters. The channel input-output relationship is defined as
\begin{equation}
Y_r(n) = \sum_{t=1}^K h_{rt}(n)X_t(n), \forall r \in \{1,2,\ldots,K\}
\end{equation}
where at channel use index $n$, $Y_r(n)$, $X_t(n)$ and $h_{rt}(n)$ are the signal observed at receiver $r$, the symbol sent from transmitter $t$ and the channel coefficient between transmitter $t$ and receiver $r$, respectively. All symbols come from a Galois field $\mathbb{GF}$ as a result of linear network coding operations at intermediate nodes inside the network over $\mathbb{GF}$.

Our interest is in minimal CSIT. Therefore, we assume the transmitters are only aware of  the  network topology information, i.e., for each channel coefficient $h_{rt}(n)$, the CSIT is only comprised of a binary variable that takes value 0 if $h_{rt}(n)=0$, and 1 if $h_{rt}(n)\neq 0$. Note that since we are interested in channel uncertainty, which is particularly relevant for larger field sizes (the binary topology knowledge would constitute perfect CSIT for  $\mathbb{GF}(2)$), we will assume throughout that the field size is large, and in particular larger than 2.  The topology information  is available globally to all transmitters and receivers. 

While we include extensions to $X$ channel and broadcast channel, our main focus is on the interference channel, where each transmitter $k$ has an independent message $W_k$ for its corresponding receiver, receiver $k$. 
Following the model in \cite{Jafar_TIM} we will assume throughout this work that the \emph{direct} channels, $h_{kk}(n)$, take only non-zero values. While this assumption is not necessary in the alternating topology case, especially when we extend the model to include $X$ channel and vector broadcast channel settings, we retain it throughout this initial exploratory work for the sake of a uniform  baseline assumption that is consistent with the original fixed topology framework of \cite{Jafar_TIM}.

\begin{figure}[h]
\begin{center}
  \includegraphics[width= 9 cm]{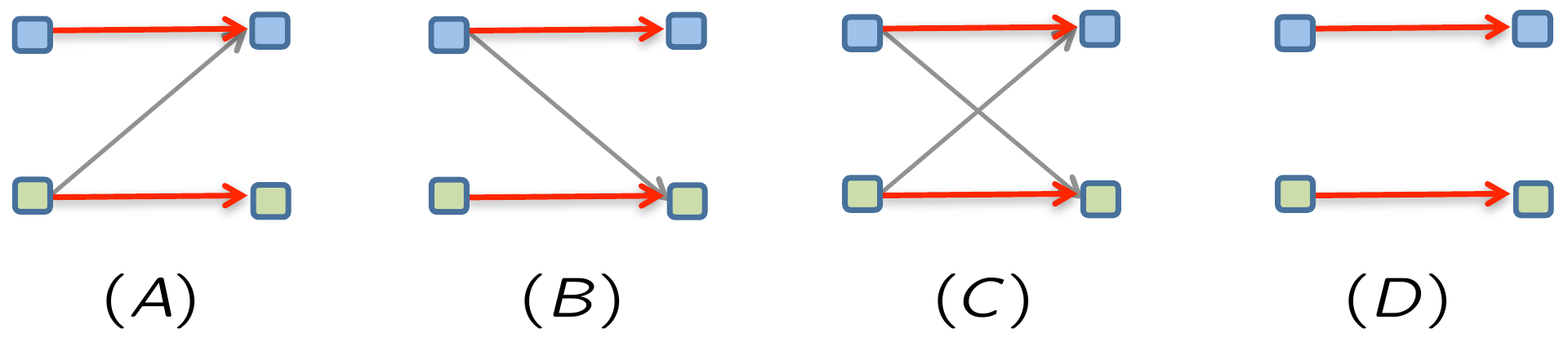}
  \caption{Two-user wired network with 4  connectivity states}
  \label{2unicast}
\end{center}
\vspace{-0.5cm}
\end{figure}

The critical aspect of this work is that the network topology is allowed to vary. For instance, the two-user wired network has 4 possible connectivity states (topologies) among which it can alternate in time, as shown in Fig. \ref{2unicast}. The fraction of time spent in each topological state is indicated with the parameter $\lambda_\cdot$, so that $\lambda_{A} + \lambda_{B} + \lambda_{C} + \lambda_{D} = 1$. The achievable rate and sum capacity are defined in the standard Shannon theoretic sense. We are interested in the capacity
of the network normalized by the capacity of a single link, i.e., a unit capacity represents $\log_2|\mathbb{GF}|$ bits/channel-use. As in \cite{Jafar_TIM} while the network itself is linear, the sources and destinations are not constrained to use linear schemes from a capacity perspective. However, the achievable schemes that we find here, several of which are shown to be capacity optimal, will turn out to be linear schemes.

\section{Main Results}
In this section, we summarize main results along with  key examples and observations.

\subsection{Two User Interference Channel}
We start with the two user interference channel. The sum capacity for this network with alternating topology is characterized in the following theorem.
\begin{figure}[h]
\begin{center}
  \includegraphics[width= 9 cm]{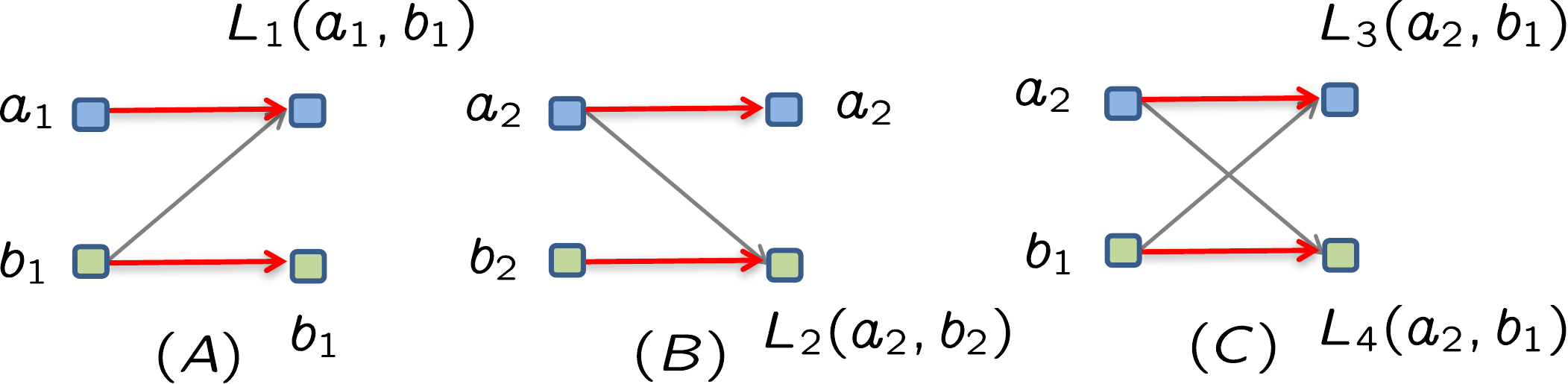}
  \caption{A 2 user interference channel with topology alternating between the three states shown. Each topology by itself has sum-capacity 1, but joint coding across the alternating topologies  achieves sum-rate 4/3. $L_i(x,y)$ represents a linear combination of $x, y$. }
  \label{zf}
\end{center}
\vspace{-0.5cm}
\end{figure}

\begin{theorem}
The sum capacity for the 2 user interference wired channel with alternating connectivity is $1+\lambda_{D}+\min(\lambda_{A},\lambda_{B},\lambda_{C})$.
\end{theorem}

The proof is given in Section IV.A. To appreciate the benefits of alternating connectivity for this network, here we illustrate the essential joint encoding scheme where sum rate $\frac{4}{3}$ is achieved by jointly coding over alternating states $A, B, C$. For this illustration assume  $(\lambda_{A},\lambda_{B},\lambda_{C},\lambda_{D}) = (\frac{1}{3}, \frac{1}{3}, \frac{1}{3},0)$ (see Fig. \ref{zf}). Note that through 3 channel uses, both receivers get 3 equations of 3 variables, 2 desired and 1 interference, from which decodability is easily seen. The key of the scheme is repeating the same interference symbol when interference link exists.  Note that in this case when we treat the states separately, the best achievable sum rate is $1$, since the sum rate is bounded by 1 in each individual state $A,B$ and $C$ \cite{Jafar_MIMO_IC}. In general, if we treat the states separately, $1+\lambda_{D}$ would be the best achievable rate. So the gain of alternating connectivity is $\min(\lambda_{A},\lambda_{B},\lambda_{C})$, which is present only when states $A,B,C$ appear simultaneously. Otherwise the topological states are separable.

\subsection{$X$ channel}
Next, we extend the model to include four independent messages $W_{rt}, r,t \in \{1,2\}$, one from each transmitter to each receiver, i.e., the two-user $X$ channel. The question we would like to explore is whether we could boost the sum capacity by adding two extra messages, compared to the interference channel. The following theorem answers this question in negative when the network is symmetric, i.e., $\lambda_{A}=\lambda_{B}$.

\begin{theorem}
The sum capacity for the symmetric two-user $X$ wired channel with alternating connectivity is $1+\lambda_{D}+\min(\lambda_{A},\lambda_{C})$.
\end{theorem}

The proof appears in Section IV.B. Similar to the interference channel, we can only achieve a rate of $1+\lambda_{D}$ at most without joint coding \cite{Jafar_MIMO_X}. 

\subsection{Vector Broadcast Channel with 2 Users}
We also consider the case with transmitter cooperation, i.e., the transmitters can exchange their messages and work in a broadcasting mode as a two-user vector broadcast channel (BC). Interestingly, the two-user vector BC with alternating connectivity is equivalent to the two-user vector BC with alternating CSIT which has been studied extensively in \cite{A-CSIT}. Specifically, the two-user wired vector BC with alternating connectivity as shown in Fig.1 is equivalent  to the two-user vector BC with alternating CSIT studied in \cite{A-CSIT} where states $A,B,C,D$ defined this work are replaced with states $(N,P), (P,N), (N,N), (P,P)$, defined in \cite{A-CSIT}, respectively. The proof is based on a  invertible transformations changing one channel model to the other, and is omitted here for brevity.

\begin{figure}[h]
\begin{center}
  \includegraphics[width= 7 cm]{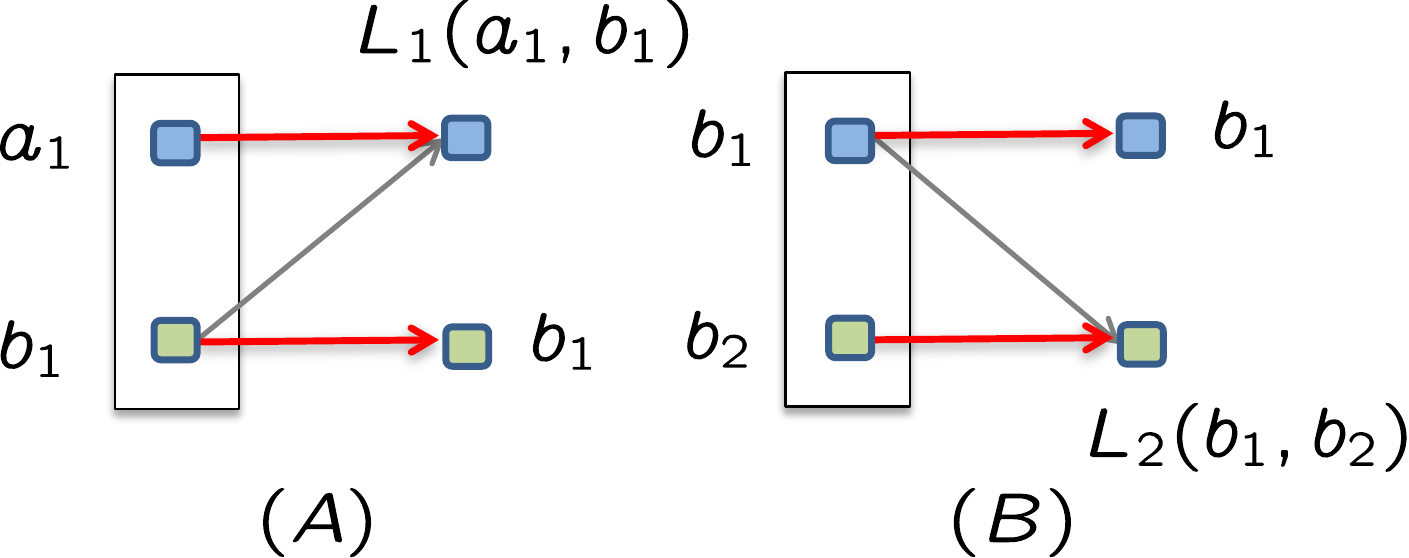}
  \caption{A 2 user vector broadcast channel with topology alternating between the two states shown. The sum-capacity of each individual topology is unknown (DoF conjectured to be 1 in the wireless case), but with alternating topologies  the sum-capacity is 3/2. Symbol $a_1$ is desired by receiver 1, and symbols $b_1,b_2$ are desired by receiver 2. All three symbols are successfully sent over two channel uses (one from each connectivity state) to achieve sum-rate 3/2.}
  \label{BC_joint_ach}
\end{center}
\vspace{-0.5cm}
\end{figure}

\begin{theorem}
The sum capacity for the symmetric two-user wired vector BC with alternating connectivity is $1+\lambda_{A}+\lambda_{D}$.
\end{theorem}

\textit{Proof:}  As stated above, we have transformed the vector BC with alternating connectivity to alternating CSIT. Now we can borrow the result shown in \cite{A-CSIT} for the DoF in wireless networks over complex field here for the capacity in wired network over finite field in normalized sense. Specifically plug in $\lambda_{PP}=\lambda_{D}$, $\lambda_{PN}=\lambda_{NP}=\lambda_{A}$ and all other state probabilities to 0 in formula (11) of Theorem 1 in \cite{A-CSIT} to produce the desired outer bound. Note that the assumption of field size being greater than 2 is important here, otherwise the network has sum-capacity equal to 2, achieved by zero-forcing.

We illustrate the key idea behind the achievable scheme in Fig. \ref{BC_joint_ach}. The symbol $a_1$ is intended for receiver $1$, and $b_1$, $b_2$ are for receiver $2$. It is easily seen that all three symbols can be decoded in two time slots. Therefore, sum rate $\frac{3}{2} (\lambda_{A} + \lambda_{B}) = 3\lambda_{A}$ is achievable for states $A$ and $B$. For states $C,D$, rate $\lambda_{C}$ and $2\lambda_{D}$ can be easily achieved. So the sum rate achieved is $3\lambda_{A} + \lambda_{C} + 2\lambda_{D} = 1+\lambda_{A}+\lambda_{D}$. \hfill $\Box$

Remarkably, without joint coding, the sum-capacity for individual states $A$ or $B$ is not known yet. In the corresponding wireless case, with channels drawn from a continuum the DoF value is conjectured to be 1 \cite{A-CSIT}, however with channels drawn from a generic set of finite cardinality (the finite state compound setting), asymptotic alignment schemes may be used to achieve the outer bound of 3/2 DoF \cite{Gou_Jafar_Wang}\cite{Ali_Compound}. Since the finite field setting naturally represents a finite state compound setting, the application of asymptotic alignment schemes to this case is an interesting research avenue. While the capacity of the individual states is not known, an immediate benefit of alternating topology is evident in the simplicity of the capacity optimal scheme in the alternating case.

\subsection{3 User Interference Channel}
\begin{figure}[h]
\begin{center}
  \includegraphics[width= 8 cm]{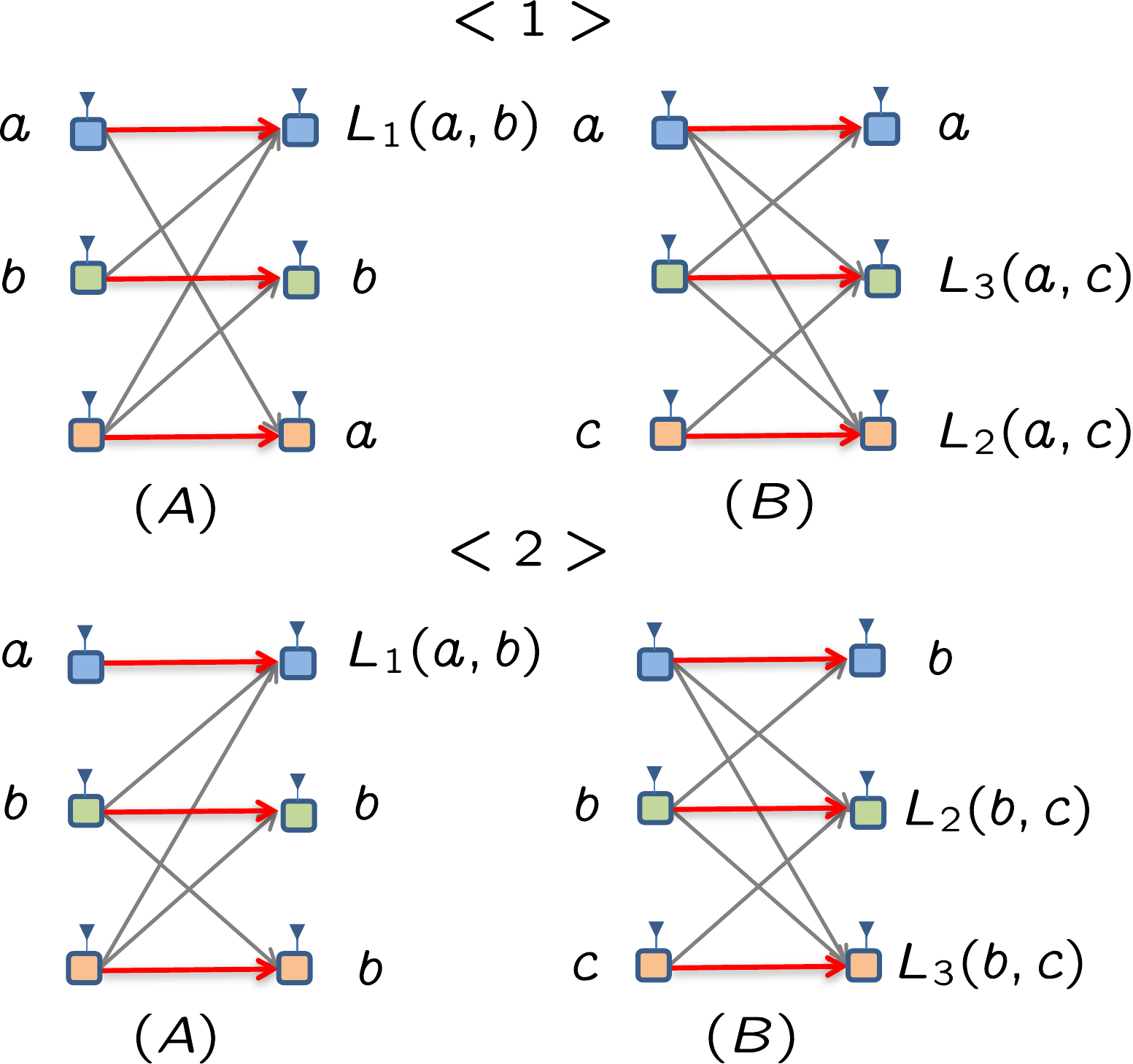}
  \caption{Two examples of the 3 user interference channel with topology alternating between the two states shown. Example 1 is at the top of the figure and example 2 is at the bottom of the figure. Each topology by itself has sum-capacity 1, but joint coding across alternating topologies achieves sum-capacity 3/2.}
  \label{3userc}
\end{center}
\vspace{-1cm}
\end{figure}

The capacity of the 3 user interference channel with alternating topology also remains open in general. However, the benefits of alternating topology are evident from the two examples shown in Fig. \ref{3userc}, where we are able to characterize capacity with and without joint coding. As it turns out in these examples,  2 states can be jointly coded to get $50\%$ gain in capacity. In particular, example 1 incorporates the idea of interference alignment, which is not needed when there are only 2 user pairs in the system. Details are given in  in Section IV.C. Also note that for 2 user interference channel, the minimum number of inseparable states is three and the gain is only $33\% (1 \rightarrow 4/3)$. These examples show the potential of bigger advantages and smarter schemes in more complex networks.

\section{Proofs}
\subsection{Proof for Theorem 1}
First, we consider outer bound. The first bound comes from the separability of the two ``Z" states, i.e., $A$ and $B$, which first appears in Theorem 7 of \cite{Jafar_EIAC}. Now consider receiver 1, we have (\ref{z1}) shown at the top of next page,
\begin{figure*}[ht]
\begin{equation}
\label{z1}
\begin{aligned}
nR_{1}\leq~& I(W_{1};Y_{1A}^n,Y_{1B}^n,Y_{1C}^n,Y_{1D}^n)+n\epsilon\\
=&H(Y_{1A}^n,Y_{1B}^n,Y_{1C}^n,Y_{1D}^n) - H(Y_{1A}^n,Y_{1B}^n,Y_{1C}^n,Y_{1D}^n | W_1) + n\epsilon \\
=&H(Y_{1A}^n,Y_{1B}^n,Y_{1C}^n,Y_{1D}^n) - H(Y_{1A}^n,Y_{1B}^n,Y_{1D}^n | W_1) -
\underbrace{H(Y_{1C}^n | W_1,Y_{1A}^n,Y_{1B}^n,Y_{1D}^n)}_{\geq 0}  + n\epsilon\\
\overset{(a)}{\leq}& H(Y_{1A}^n,Y_{1B}^n,Y_{1C}^n,Y_{1D}^n) - H(Y_{1A}^n,Y_{1B}^n,Y_{1D}^n | W_1,X_{1A}^n,X_{1B}^n,X_{1C}^n,X_{1D}^n) + n\epsilon\\
=& H(Y_{1A}^n,Y_{1B}^n,Y_{1C}^n,Y_{1D}^n) - H(h_{11A}^nX_{1A}^n+h_{12A}^nX_{2A}^n,h_{11B}^nX_{1B}^n,h_{11D}^nX_{1D}^n | W_1,X_{1A}^n,X_{1B}^n,X_{1C}^n,X_{1D}^n)+ n\epsilon\\
=& H(Y_{1A}^n,Y_{1B}^n,Y_{1C}^n,Y_{1D}^n) - H(h_{12A}^nX_{2A}^n | W_1,X_{1A}^n,X_{1B}^n,X_{1C}^n,X_{1D}^n)+ n\epsilon\\
\overset{(b)}{=}& H(Y_{1A}^n,Y_{1B}^n,Y_{1C}^n,Y_{1D}^n) - H(h_{22A}^nX_{2A}^n)+ n\epsilon\\
\leq&H(Y_{1A}^n) + H(Y_{1B}^n) + H(Y_{1C}^n) + H(Y_{1D}^n) - H(Y_{2A}^n) + n\epsilon
\end{aligned}
\end{equation}
\hrulefill
\end{figure*}
where (a) follows from the observation that ($X_{1A}^n,X_{1B}^n,X_{1C}^n,X_{1D}^n$) is a function of $W_1$, (b) is due to the fact that $X_{2A}^n$ is independent of $W_1$ and scaling does not influence entropy. Similarly, for receiver 2, we have
\begin{equation}
\label{z2}
nR_{2}\leq H(Y_{2A}^n) + H(Y_{2B}^n) + H(Y_{2C}^n) + H(Y_{2D}^n) - H(Y_{1B}^n) + n\epsilon.
\end{equation}
Adding (\ref{z1}) and (\ref{z2}), we have
\begin{equation}
\begin{aligned}
n(R_{1}+R_{2})\leq~&H(Y_{1A}^n)+H(Y_{2B}^n)+H(Y_{1C}^n)\\
&+H(Y_{2C}^n)+H(Y_{1D}^n)+H(Y_{2D}^n)+n\epsilon\\
\overset{(a)}{\leq}&(\lambda_{A}+\lambda_{B}+2\lambda_{C}+2\lambda_{D})n + n\epsilon\\
=&(1+\lambda_{C}+\lambda_{D})n + n\epsilon
\end{aligned}
\end{equation}
where (a) follows from the fact that all random variables come from $\mathbb{GF}$ and the last step follows from $\lambda_{A} + \lambda_{B} + \lambda_{C} + \lambda_{D} = 1$. Normalizing by $n$, we have
\begin{equation}
\label{z}
R_{1}+R_{2} \leq 1+\lambda_{C}+\lambda_{D}.
\end{equation}
The second bound comes from genie-aided MAC bound. We provide genie $Y_{2B}^n$ and $Y_{2D}^n$ to receiver $1$. Then receiver $1$ can reconstruct the received signal at receiver 2 at state $B$ and $D$. Also, at state $A$ and $C$, after decoding and subtracting its desired signal, receiver 1 can also get the exact output at receiver 2. Now receiver 1 can decode $W_{2}$ if receiver 2 can, which is true by system design. So receiver 1 can decode both $W_1$ and $W_2$. We have
\begin{equation}
\begin{aligned}
&n(R_{1}+R_{2})\\
\leq &I(W_{1},W_{2};Y_{1}^n,Y_{2B}^n,Y_{2D}^n)+n\epsilon\\
=&H(Y_{1}^n,Y_{2B}^n,Y_{2D}^n) - \underbrace{H(Y_{1}^n,Y_{2B}^n,Y_{2D}^n|W_{1},W_{2})}_{=0} +n\epsilon\\
\leq&H(Y_{1}^n)+H(Y_{2B}^n)+H(Y_{2D}^n) +n\epsilon\\
\leq&(1+\lambda_{B}+\lambda_{D})n+n\epsilon.
\end{aligned}
\end{equation}
Dividing by $n$, we have
\begin{equation}
\label{mac}
\begin{aligned}
R_{1}+R_{2} \leq 1+\lambda_{B}+\lambda_{D}.
\end{aligned}
\end{equation}
Symmetrically, we can also provide the $Y_{1A}^n$ and $Y_{1D}^n$ to receiver $2$ so that it can decode both messages $W_{1}$ and $W_{2}$. Then we have, similarly,
\begin{equation}
\label{mac2}
R_{1}+R_{2} \leq 1+\lambda_{A}+\lambda_{D}.
\end{equation}
Combining (\ref{z}), (\ref{mac}) and (\ref{mac2}), we prove the outer bound.

Next, we proceed to the achievable scheme. Based on the constituent encoding scheme given in Section III, where sum rate $\frac{4}{3}$ is achieved when states $A$, $B$, and $C$ occur with equal probability $\frac{1}{3}$, we can obtain the general achievable scheme. In short, besides joint coding over equal-fraction of states $A,B,C$, we treat all the other remaining states separately.
Note that the capacity outer bound takes two different forms, depending on whether $\lambda_{C}$ is larger than $\min(\lambda_{A},\lambda_{B})$ or not. Thus we consider the following two mutually exclusive cases:

\begin{itemize}
  \item {Case A}: When $\lambda_{C} > \min(\lambda_{A},\lambda_{B})$, the outer bound is $1+\min(\lambda_{A},\lambda_{B})+\lambda_{D}$. We can group states $A,B,C$ at a fraction of $\min(\lambda_{A},\lambda_{B})$. For interference-free state $D$, we would get rate 2 naturally. Time sharing is used in all remaining cases. The total sum rate achieved is
  \begin{equation}
  \begin{aligned}
  R_{1}+R_{2} =& 3\min(\lambda_{A},\lambda_{B})  \times 4/3 + \lambda_{D} \times 2 \\
&+ [1-3\min(\lambda_{A},\lambda_{B})-\lambda_{D}] \times 1\\
=& 1+\min(\lambda_{A},\lambda_{B})+\lambda_{D}.
  \end{aligned}
  \end{equation}

  \item {Case B}: When $\lambda_{C} \leq \min(\lambda_{A},\lambda_{B})$, the outer bound is $1+\lambda_{C}+\lambda_{D}$. Similarly, we group states $A,B,C$ as much as possible. As desired, the total achievable rate is
  \begin{equation}
  \begin{aligned}
  R_{1}+R_{2} &= 3\lambda_{C} \times 4/3 + \lambda_{D} \times 2 + (1-3\lambda_{C}-\lambda_{D}) \times 1\\
&= 1 + \lambda_{C}+\lambda_{D}.
  \end{aligned}
  \end{equation}

\end{itemize}

\subsection{Proof for Theorem 2}
As the capacity is already achievable in the two unicast setting, we can employ the same achievable scheme as shown in Section IV.A by setting $W_{12},W_{21} = \phi$. Therefore, here we only need to prove the outer bound. The key step for the first bound is to prove states $A$ and $B$ are separable. We give $W_{21}$ as genie to receiver 1, then from Fano's inequality we can obtain (\ref{x1}) (see next page),
\begin{figure*}[t]
\begin{equation}
\label{x1}
\begin{aligned}
n(R_{11}+R_{12})\leq&I(W_{11},W_{12};Y_{1A}^n,Y_{1B}^n,Y_{1C}^n,Y_{1D}^n,W_{21})+n\epsilon\\
\overset{(a)}{=}&I(W_{11},W_{12};Y_{1A}^n,Y_{1B}^n,Y_{1C}^n,Y_{1D}^n|W_{21})+n\epsilon\\
=& H(Y_{1A}^n,Y_{1B}^n,Y_{1C}^n,Y_{1D}^n|W_{21}) - H(Y_{1A}^n,Y_{1B}^n,Y_{1C}^n,Y_{1D}^n|W_{21},W_{11},W_{12}) +n\epsilon\\
=& H(Y_{1A}^n,Y_{1B}^n,Y_{1C}^n,Y_{1D}^n|W_{21}) - H(Y_{1A}^n|W_{21},W_{11},W_{12}) - \underbrace{H(Y_{1B}^n,Y_{1C}^n,Y_{1D}^n|Y_{1A}^n,W_{21},W_{11},W_{12})}_{\geq 0}+n\epsilon \\
\overset{(b)}{\leq}& H(Y_{1A}^n,Y_{1B}^n,Y_{1C}^n,Y_{1D}^n|W_{21}) - H(X_{2A}^n|W_{21},W_{11},W_{12}) + n\epsilon \\
\overset{(c)}{=}& H(Y_{1A}^n,Y_{1B}^n,Y_{1C}^n,Y_{1D}^n|W_{21}) - H(X_{2A}^n|W_{12}) + n\epsilon \\
\leq& H(Y_{1A}^n|W_{21})+H(Y_{1B}^n|W_{21})+H(Y_{1C}^n|W_{21})+H(Y_{1D}^n|W_{21}) - H(Y_{2A}^n|W_{12}) + n\epsilon \\
\end{aligned}
\end{equation}
\hrulefill
\end{figure*}
where (a) follows from the messages are independent, (b) is due to fact that after knowing $W_{11}$ and $W_{21}$, receiver 1 can reconstruct and subtract the signal $X_{1A}^n$ from the received signal to leave $X_{2A}^n$ only and (c) follows from the transmitted signal $X_{2A}^n$, which originates from transmitter 2, is independent of $W_{11}$ and $W_{21}$, messages from transmitter 1.
Similarly, for receiver $2$, we have
\begin{equation}
\begin{aligned}
\label{x2}
&n(R_{21}+R_{22})\\
\leq& H(Y_{2A}^n|W_{12})+ H(Y_{2B}^n|W_{12})+  H(Y_{2C}^n|W_{12})\\ 
& +  H(Y_{2D}^n|W_{12}) -H(Y_{1B}^n|W_{21})+n\epsilon.
\end{aligned}
\end{equation}
Adding (\ref{x1}) and (\ref{x2}), we have
\begin{equation}
\begin{aligned}
&n(R_{11}+R_{12}+R_{21}+R_{22})\\
\leq& H(Y_{1A}^n|W_{21})+H(Y_{1C}^n|W_{21})+H(Y_{1D}^n|W_{21})\\
&+H(Y_{2B}^n|W_{12})+  H(Y_{2C}^n|W_{12}) +  H(Y_{2D}^n|W_{12}) +n\epsilon\\
\leq& (\lambda_A + \lambda_B + 2\lambda_C + 2\lambda_D)n + n\epsilon\\
=& (1 + \lambda_C + \lambda_D)n  + n\epsilon.
\end{aligned}
\end{equation}
Normalizing by $n$, we get the first sum rate bound
\begin{equation}
\label{x}
R_{11}+R_{12}+R_{21}+R_{22} \leq 1 + \lambda_C + \lambda_D.
\end{equation}
The second sum rate outer bound comes from two-user vector BC since transmitters cooperation can not reduce capacity. According to Theorem 3, we have
\begin{equation}
R_{11}+R_{12}+R_{21}+R_{22} \leq 1 + \lambda_A + \lambda_D,
\end{equation}
which completes the proof.

\subsection{3 User Interference Channel}

For the network topology shown in Fig. 4, we wish to show that these two states have capacity 1 individually while capacity increases to 3/2  jointly. 
We prove example 1 and example 2 follows similarly. First, let us look at the states separately. Rate 1 can be achieved easily. The outer bound can be seen as follows. We only prove state ($A$) and state ($B$) follows similarly. As the transmitter only knows the connectivity, we can assume the value of connected links are all 1 without reducing the capacity region. Now consider receiver 1, we argue that it can decode all three messages. Since desired message is decodable by design, receiver 1 can reliably reconstruct and subtract it from its received signal. This gives receiver 1 exactly the same output as observed by receiver 2. As receiver 2 can decode its desired message, so can receiver 1. Now receiver 1 can subtract the effect of $W_2$ from its received signal, which gives it the exact same signal seen by receiver 3. So receiver 1 can also decode $W_3$ if receiver 3 can, which is satisfied by any coding scheme. So the sum capacity of this network is bounded by the sum-capacity of the MAC to receiver 1, which is 1. Next, we consider the two states jointly. The achievable scheme is shown in the Fig. \ref{3userc}. Receiver 1 and 3 both have two equations formed by two variables from which they can decode successfully. Receiver 2 can also resolve its desired symbol $b$ since $b$ is received interference free over state ($A$). Interestingly, its interference $a$ and $c$ align at state ($B$). As 3 symbols are transmitted over 2 states, we achieve rate 3/2 in total. Lastly, we establish the optimality of this result. We wish to show the sum rate of any two messages can not exceed 1. Consider any two user pairs, for example, 1 and 2. Essentially we eliminate $W_3$ by giving it as genie everywhere which can not hurt the rate of $W_1$ and $W_2$. Now we apply Theorem 1 where $\lambda_{A} = 1$ and all the other terms are zero. Then $R_1 + R_2 \leq 1$ and by repeating this argument for each two user pairs, we conclude that the sum capacity of this network is 3/2.

\section{Conclusion}
The implications of alternating topology are explored for the topological interference management problem. The  extension is highly non-trivial since it goes beyond the index coding setting and reveals the synergistic benefits of alternating connectivity.

\end{document}